\documentclass[a4paper,fleqn,usenatbib]{mnras}

\usepackage{newtxtext,newtxmath}
\usepackage[T1]{fontenc}
\usepackage{ae,aecompl}

\usepackage{graphicx}	% Including figure files
\usepackage{amsmath}	% Advanced maths commands
\usepackage{amssymb}	% Extra maths symbols
\usepackage{multirow}

\title[Posterior Propriety for Proper Bayesian Analyses]{How proper are  Bayesian models in the astronomical literature?}

\author[Tak et al.]{
Hyungsuk Tak,$^{1}$\thanks{E-mail: hyungsuk.tak@gmail.com}
Sujit K. Ghosh,$^{2}$
and Justin A. Ellis$^{3}$
\\
$^{1}$Department of Applied and Computational Mathematics and Statistics, University of Notre Dame, Notre Dame, IN 46556, USA\\
$^{2}$Department of Statistics, North Carolina State University, Raleigh, NC 27695, USA\\
$^{3}$Infinia ML, Durham, NC 27701, USA
}

\date{Accepted XXX. Received YYY; in original form ZZZ}

\pubyear{2015}

\begin{document}
\label{firstpage}
\pagerange{\pageref{firstpage}--\pageref{lastpage}}
\maketitle

\begin{abstract}
The well-known Bayes theorem assumes that a  posterior distribution  is a  probability distribution.  However, the posterior distribution may no longer be a  probability distribution if an improper prior distribution (non-probability measure) such as an unbounded uniform prior is used. Improper priors are often used in the astronomical literature to reflect  a lack of prior knowledge, but checking whether the resulting posterior is a probability distribution is sometimes neglected. It turns out that \textcolor{black}{23} articles out of 75 articles  (\textcolor{black}{30.7}\%) published online in two renowned astronomy journals (\emph{ApJ} and \emph{MNRAS}) between Jan 1, 2017 and Oct 15, 2017 make use of Bayesian analyses without rigorously establishing posterior propriety. A disturbing aspect is that a Gibbs-type Markov chain Monte Carlo (MCMC) method can produce a seemingly reasonable posterior sample even when the posterior is not a probability distribution \citep{hobert1996propriety}. In such cases, researchers may erroneously make probabilistic inferences without noticing that the MCMC sample is  from a non-\textcolor{black}{existing} probability distribution.  We review why checking  posterior propriety  is fundamental in Bayesian analyses\textcolor{black}{,} and discuss how \textcolor{black}{to}  set up scientifically motivated proper priors.
\end{abstract}

%  \textcolor{black}{to} avoid the pitfalls of using improper priors
%\textcolor{black}{for improper priors}

%%%%%%%%%%%%%%%%%%%%%%%%%%%%%%%%%%%%%%%%%%%%%%%%%%

%%%%%%%%%%%%%%%%% BODY OF PAPER %%%%%%%%%%%%%%%%%%

\section{Introduction}\label{sec:intro}

%resulting
A Bayesian model is uniquely determined by two components: (i) a likelihood function  of unknown parameters $\boldsymbol{\theta}$ given the data $\boldsymbol{y}$ denoted by $L(\boldsymbol{\theta}; \boldsymbol{y})$, which is proportional to a conditional probability density $f(\boldsymbol{y}\mid\boldsymbol{\theta})$ of a sampling distribution,  and (ii) a joint prior density, $p(\boldsymbol{\theta})$. Using the fundamental Bayes theorem \textcolor{black}{(see Appendix~\ref{appendixBayes} for  details)}, we can derive the  posterior density of $\boldsymbol{\theta}$ as follows\footnote{Within a finite-dimensional parametric framework, all density functions are formally defined with respect to a common dominating $\sigma$-finite measure like Lebesgue measure (or counting measure).}:
\begin{equation}\label{bayes}
\pi(\boldsymbol{\theta}\mid \boldsymbol{y})=\frac{f(\boldsymbol{y}\mid\boldsymbol{\theta})p(\boldsymbol{\theta})}{\int f(\boldsymbol{y}\mid\boldsymbol{\theta})p(\boldsymbol{\theta})d\boldsymbol{\theta}}= \frac{L(\boldsymbol{\theta}; \boldsymbol{y})p(\boldsymbol{\theta})}{\int L(\boldsymbol{\theta}; \boldsymbol{y})p(\boldsymbol{\theta})d\boldsymbol{\theta}}.
\end{equation}
 Even if \textcolor{black}{the} joint prior  is improper (i.e., $\int p(\boldsymbol{\theta})d\boldsymbol{\theta}=\infty$), the posterior density in Equation~\eqref{bayes} can still be a valid probability density as long as the denominator is finite given the data $\boldsymbol{y}$, i.e., $\int L(\boldsymbol{\theta}; \boldsymbol{y})p(\boldsymbol{\theta})d\boldsymbol{\theta}<\infty$. The finite integrability of the product $L(\boldsymbol{\theta};\boldsymbol{y})p(\boldsymbol{\theta})$  is called {\em posterior propriety}. It is often  unnecessary to compute \textcolor{black}{this integral because only a posterior kernel function,  $q(\boldsymbol{\theta}\mid \boldsymbol{y})\equiv L(\boldsymbol{\theta}; \boldsymbol{y})p(\boldsymbol{\theta}),$ which is proportional to $\pi(\boldsymbol{\theta}\mid \boldsymbol{y})$  if posterior propriety holds, is required to implement most MCMC algorithms}. 
\\
\indent Posterior propriety \textcolor{black}{is crucial} in MCMC\textcolor{black}{, ensuring a couple of conditions for the convergence of a Markov chain. An irreducible, aperiodic, and recurrent Markov chain converges to a unique stationary distribution, and posterior propriety (with a random walk proposal) guarantees the aperiodicity and recurrence \citep[p.~279,][]{gelman2013bayesian, tierney1994}}. 
\\
\indent However, posterior propriety \textcolor{black}{does} not necessarily hold if the prior $p$ is improper. For example,  uniform($0, \infty$) and uniform($-\infty, \infty$)  are widely used improper priors. \textcolor{black}{Adopting such improper priors, one may fail to check posterior propriety because most MCMC methods do not require users to check posterior propriety, i.e., $\int L(\boldsymbol{\theta}; \boldsymbol{y})p(\boldsymbol{\theta})d\boldsymbol{\theta}<\infty$.} When the posterior is improper, the most serious issue is that a Gibbs-type MCMC method may still appear to work well by producing a seemingly reasonable posterior sample from the path of the Markov chain \citep{hobert1996propriety}. Consequently, \textcolor{black}{researchers} may continue  making \textcolor{black}{posterior} inferences without knowing that the \textcolor{black}{MCMC} sample is in fact drawn from a non-existent posterior \textcolor{black}{probability} distribution.  \cite{hobert1996propriety} first warned about this insidious feature of posterior impropriety. \textcolor{black}{To prevent this, they recommended} either proving posterior propriety (analytically) for improper priors  or using jointly proper priors. Since then, statisticians have rigorously established posterior propriety using analytical techniques when improper priors are employed \citep{daniels1999prior, natarajan2000reference, tak2017data}.
%before using  \textcolor{black}{MCMC} methods 
%using these posterior sample 
%the  posterior distribution is not a probability distribution and that 

Posterior propriety is sometimes neglected in the astronomical literature. Our investigation reveals that  \textcolor{black}{23} articles out of 75 (\textcolor{black}{$30.7$}\%) published online in \emph{ApJ} and \emph{MNRAS} between Jan 1, 2017 and Oct 15, 2017 report Bayesian analyses without rigorously establishing  posterior propriety.  We hope that the posterior distributions of these 24 articles are actually proper, although it remains an open issue until  posterior propriety is analytically established. 
%that needs to be addressed before any scientific conclusions are made based on the MCMC samples.
% because its proof is the only way to completely exclude a possibility that their posterior samples are drawn from non-existent posterior distributions.

The rest of this article is organized as follows.  Section~\ref{section2example} introduces a simple but non-trivial example of using an MCMC method \textcolor{black}{for} an improper posterior distribution. In Section~\ref{section3propriety}, we \textcolor{black}{investigate posterior propriety in 75 articles published online in \emph{ApJ} and \emph{MNRAS}.} Section~\ref{section4solution} discusses several ways to prove posterior propriety, focusing on using scientifically motivated proper priors \textcolor{black}{which automatically guarantees} posterior propriety.

\section{A Simple but Non-trivial Example}\label{section2example}

Here we reproduce a classical example of \cite{hobert1996propriety} that handles a Gaussian hierarchical model commonly used in Bayesian analyses. Suppose the observation $y_j$ ($j=1, \ldots, n$) follows an independent Gaussian distribution given unknown mean $\mu_j$ with known  measurement variance $V_j$. \textcolor{black}{Also,} $\mu_j$  follows another independent Gaussian distribution with unknown mean $\theta$ and unknown variance $\sigma^2$:
\begin{equation}\label{normalmodel}
y_j \mid \mu_j \sim \textrm{N}(\mu_j, V_j)~~\textrm{and}~~\mu_j \mid \theta, \sigma^2 \sim \textrm{N}(\theta, \sigma^2).
\end{equation}
We set  up a joint prior kernel function of $\theta$ and $\sigma^2$ as 
\begin{equation}\label{improperprior}
p_1(\theta, \sigma^2)=p_1(\theta)p_1(\sigma^2)\propto\frac{1}{\sigma^2},
\end{equation}
which is  improper because $\int_0^{\infty}\int_{-\infty}^{\infty} p_1(\theta, \sigma^2) d\theta~\!\! d\sigma^2=\infty$. The prior on $\sigma^2$ in Equation~\eqref{improperprior} is equivalent to both $d\sigma/\sigma$ and $d\log(\sigma)$, i.e., a \textcolor{black}{widely} used \textcolor{black}{improper} flat prior on a logarithmic scale of $\sigma$. The resulting posterior kernel function  is
\begin{equation}\label{improperpost}
q(\boldsymbol{\mu}, \theta, \sigma^2 \mid \boldsymbol{y}) = p_1(\theta, \sigma^2)\prod_{j=1}^{n}\left[f(y_j\mid \mu_j)~ p(\mu_j\mid \theta, \sigma^2)\right],
\end{equation}
where $\boldsymbol{\mu}=(\mu_1, \ldots, \mu_n)$, $\boldsymbol{y}=(y_1, \ldots, y_n)$, and density functions $f$ and $p$ are defined by Equation~\eqref{normalmodel}.  This  posterior kernel function  is improper  due to the prior on $\sigma^2$ regardless of the data; see Appendix~\ref{appendixB} for a proof.  % Sujit, this parenthesis is to make Kaisey feel less awkward because his article uses an improper flat on a Gaussian mean.
% and \eqref{improperprior}, respectively

%To sample the posterior in equation~\eqref{improperpost}, w
Although the posterior kernel function in Equation~\eqref{improperpost} is not a probability density, we can still derive  its MCMC sampling scheme. \textcolor{black}{Following \cite{hobert1996propriety}, we set $\boldsymbol{y}=(-10, 10)$, $n=2$, and $V_j=1$},  but we keep using the notation $V_j$, $y_j$, and $n$ for generality. We use a Gibbs sampler \citep{geman1984stochastic} that iteratively samples the following  conditional posterior distributions: For $j=1, 2$,
\begin{align}
\begin{aligned}\label{conditionals}
&\mu_j\mid \boldsymbol{\mu}_{[-j]}, \theta, \sigma^2,  \boldsymbol{y}  \sim\textrm{N}\left(~\frac{\sigma^2y_j+V_j\theta}{V_j + \sigma^2},~ \frac{\sigma^2}{V_j + \sigma^2}~\right),\\
&\theta\mid \boldsymbol{\mu}, \sigma^2,  \boldsymbol{y} \sim \textrm{N}\left(~\bar{\mu},~ \frac{\sigma^2}{n}~\right),\\
&\sigma^2\mid \boldsymbol{\mu}, \theta,  \boldsymbol{y}  \sim \textrm{in}\textrm{verse-Gamma}\left(~\frac{n}{2},~ \frac{\sum_{j=1}^n(\mu_j-\theta)^2}{2}~\right),
\end{aligned}
\end{align}
where $\boldsymbol{\mu}_{[-j]}$ denotes $\boldsymbol{\mu}$ without the $j$th component, $\bar{\mu}$ \textcolor{black}{is the average of the elements of  $\boldsymbol{\mu}$}, and the inverse-Gamma($a, b$) kernel function of $x$ is $x^{-a-1}\exp(-b/x)$. At iteration $i$, for example, this \textcolor{black}{Gibbs} sampler updates each parameter in a sequence, i.e., $(\boldsymbol{\mu}^{(i)}, \theta^{(i-1)}, \sigma^{2(i-1)})$, $(\boldsymbol{\mu}^{(i)}, \theta^{(i)}, \sigma^{2(i-1)})$, and $(\boldsymbol{\mu}^{(i)}, \theta^{(i)}, \sigma^{2(i)})$.
 Almost all MCMC schemes for sampling \textcolor{black}{multiple} parameters use such Gibbs-type  updates (either \textcolor{black}{single-coordinate-wise} or block-wise) at each iteration to form a Markov chain. We set the initial values as $\boldsymbol{\mu}^{(0)}=(-10, 10)$, $\theta^{(0)}= 0$, and $\sigma^{2(0)}=1$, and draw 10,000 posterior samples of each parameter.

\begin{figure*}
\begin{center}
\includegraphics[scale = 0.5]{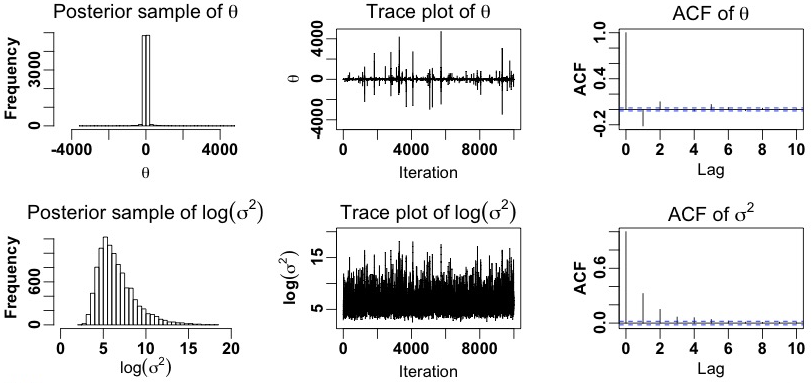}
\caption{The result of sampling an improper posterior distribution in Equation~\eqref{improperpost}. The histogram, trace plot, and auto-correlation function of 10,000 posterior samples of $\theta$ are on the top  panels and those of $\log(\sigma^2)$ on the bottom panels. The Markov chain appears to converge  to a certain probability distribution, although the target posterior distribution in Equation~\eqref{improperpost} is not a probability distribution.\label{ex1figure}}
\end{center}
\end{figure*}

%draw 10,000 posterior samples of each unknown parameter and 
In Figure~\ref{ex1figure}, we display the histogram, trace plot, and auto-correlation function  of 10,000 posterior samples of $\theta$ on the top  and those of $\log(\sigma^2)$ on the bottom. The posterior sample of $\theta$ concentrates on zero and that of $\log(\sigma^2)$ also forms a unimodal histogram. The trace plots show that the Markov chain explores the parameter space rapidly and the auto-correlation functions decrease quickly. The effective sample size\footnote{\textcolor{black}{The effective sample size is defined as $n/(1 + 2\sum_{i=1}^{\infty}\rho(i))$, where $n$ is the length of a Markov chain and $\rho(i)$ is the auto-correlation at lag $i$. The effective sample size becomes $n$ if the sample from the path of a Markov chain is independent, i.e., $\rho(i)=0$ for all $i$. We use a function \texttt{effectiveSize} of an R package \texttt{coda} \citep{plummer2006coda} to estimate the effective sample size.}} of $\theta$ is \textcolor{black}{8,140} and that of $\log(\sigma^2)$ is \textcolor{black}{1,847}. Clearly, the Markov chain appears to converge to a certain probability distribution, and thus it makes sense to make a probabilistic inference using this posterior sample. However, if the initial value of $\sigma^2$ were close to zero at which the posterior kernel function puts infinite mass, the Markov chain would stay at $\sigma^2=0$ permanently without producing such a seemingly reasonable posterior sample. \textcolor{black}{See} \cite{hobert1996propriety} for more theoretical details.
%Although the sample comes from a non-existent probability distribution, there is no  evidence against the convergence of the Markov chain.
% shown in Figure~\ref{ex1figure}
%unless we  know that , 
%The reason  this MCMC  scheme produced a seemingly reasonable posterior sample despite posterior impropriety is that the Markov chain did not begin with the initial values of $\sigma^{2}$ close to zero . 

Such an inappropriate probabilistic inference based on a non-\textcolor{black}{existing} probability distribution can actually happen in reality unless posterior propriety is proven in advance. The article of \cite{pihajoki2017a} published in \emph{MNRAS} uses a similar but more  complicated  Gaussian hierarchical model that can be built upon a marginalized model of Equation~\eqref{normalmodel}, that is,
\begin{equation}\label{marginaldata}
y_j \mid \theta, \sigma^2 \sim \textrm{N}(\theta, V_j+\sigma^2).
\end{equation}
\textcolor{black}{A model of} \cite{pihajoki2017a} replaces $\theta$ in Equation~\eqref{marginaldata} with $\alpha+\beta x_j$, where $\alpha$ and $\beta$ are unknown regression coefficients and $x_j$ is some known covariate information with its known measurement variance $V_{x_j}$. Also, \textcolor{black}{the model} replaces $V_j$ in Equation~\eqref{marginaldata} with $\beta^2V_{x_j}+V_j-2\beta\rho (V_{x_j}V_j)^{0.5}$, multiplies $\sigma^2$ in Equation~\eqref{marginaldata} by \textcolor{black}{$1+\beta^2$}, and adopts an improper \textcolor{black}{joint} prior $\textcolor{black}{d\alpha} d\sigma/\sigma$; see Equations (33)--(37) of \cite{pihajoki2017a}. \textcolor{black}{This improper joint prior} is equivalent to the problematic choice in Equation~\eqref{improperprior}. The resulting posterior is not a probability distribution. This is because when $\beta=0$, the model of \cite{pihajoki2017a} becomes exactly the same as the one in Equation~\eqref{marginaldata} that is improper with $\textcolor{black}{d\alpha} d\sigma/\sigma$. Therefore, the integral of the posterior kernel function of \cite{pihajoki2017a} is not finite.
\\
\indent\textcolor{black}{The article of \cite{pihajoki2017a} does not check posterior propriety before using an MCMC method. Thus, without recognizing posterior impropriety,} the article makes a probabilistic inference using the seemingly reasonable posterior sample drawn from a non-existent posterior  distribution.  An MCMC method for this model may not show \textcolor{black}{any} evidence of posterior impropriety unless a Markov chain starts with the initial value of $\sigma^2$ close to zero.  In practice, \textcolor{black}{however,} the inference \textcolor{black}{in \cite{pihajoki2017a}  may be similar to \textcolor{black}{that} based on a proper posterior equipped with weakly informative proper priors. This is} because it is likely that the \textcolor{black}{Markov chain} of \cite{pihajoki2017a} resides in a safe (high-likelihood) region  without exploring the entire parameter space.
\\
\indent\textcolor{black}{One may think that we are exaggerating a problem with a pathological example where a tiny corner of the parameter space becomes a problem. We emphasize  again that our concern is whether researchers are clearly aware that their Bayesian inferences are based on \emph{probability distributions}. We have used such a tiny parameter space, e.g, $\sigma^2\in[0, \epsilon)$ and $\beta=0$,  to raise a question about this concern, not to criticize \cite{pihajoki2017a}'s omission in exploring this  pathological region. Exploring the entire parameter space, however,  is a useful practice to check a Markov chain's convergence. Inconsistent results from multiple Markov chains, whose initial values are spread across the parameter space, indicate the lack of convergence, e.g., due to multimodality or possibly posterior impropriety. A popular convergence diagnostic statistic of Gelman and Rubin (1992) is based on this idea.  Initiating multiple Markov chains at least one of which begins near $\sigma^2=0$ might have indicated posterior impropriety in the case of \cite{pihajoki2017a}.}
\\
\indent \textcolor{black}{An astronomer's intuition or prior knowledge may  indicate which parameter space is scientifically meaningful to search a priori. This is invaluable information, but should be used  carefully because one may have an incentive to initiate a Markov chain only in such a specific part of the parameter space. This chain might have stayed in that part, inevitably producing a  result that is consistent with the astronomer's intuition. But, it is not desirable to report this result as if the entire parameter space were  explored (even though a physically inspired model may have more power to constrain the region of interest than a non-physically inspired ones). Without being fully informed of such a limited search, readers may assume that evidence for multiple modes or posterior impropriety has not been found in the entire parameter space. Therefore, it is desirable to run multiple Markov chains with widely spread initial values across the  parameter space or to use more tightly bounded  priors to clarify which part is actually explored.}

\section{Posterior propriety in the astronomical literature}\label{section3propriety}
We investigated the literature published online in \emph{ApJ} and \emph{MNRAS} between Jan~1, 2017 and Oct~15, 2017.  \textcolor{black}{On the webpages of IOPscience\footnote{http://iopscience.iop.org/} and \emph{MNRAS}\footnote{https://academic.oup.com/mnras}, we found 75 articles} whose titles or abstracts contain a word `Bayesian'\textcolor{black}{; see Appendix~\ref{appendixC} for details of the selection}. None of the  75 articles mention posterior propriety, and thus we \textcolor{black}{checked} further by classifying them into three categories; (a) priors are jointly proper; (b) priors are jointly improper; and (c) priors are not clearly specified. The last category includes cases where uniform (or flat) prior distributions are used without clearly specified ranges. Table~\ref{table:literature} summarizes the classification; \textcolor{black}{also} see Appendix~\ref{appendixC} for details. More than half of the articles use jointly proper priors. \textcolor{black}{However,} there are \textcolor{black}{23} articles in categories~(b) and (c) that need  proofs for posterior propriety to assure that their scientific arguments are actually based on \textcolor{black}{proper} posterior distributions. 
%remaining
%Bayesian methods are developed in other articles but are used without details. We also classify  articles into category (c) if they adopt 

%no one can ever reproduce their scientific results 
%knowing
The issue of the 20 articles in category (c) is not only posterior propriety but also reproducibility because \textcolor{black}{their results cannot be reproduced without  information about their priors.} \textcolor{black}{For instance, there are infinitely many uniform prior distributions according to their ranges, and thus a flat uniform prior is not a clear description.} Proving posterior propriety  can contribute to reproducible science as a  by-product because  its first step is to \textcolor{black}{specify} a Bayesian model \textcolor{black}{clearly, i.e., a} likelihood function of unknown parameters and their prior distributions.

\textcolor{black}{Without hurting readability, one may be able to specify both likelihood function and priors in an appendix, mentioning only the resulting posterior propriety  in the main text. This practice will greatly improve statistical clarity and reproducibility in the  astronomical literature.}
%A Bayesian model is uniquely determined once the likelihood function and prior distributions are  specified. 
%But it is impossible to judge posterior propriety in 

\begin{table}
	\centering
	\caption{Classification of 75 articles published online in $ApJ$ and $MNRAS$ between Jan 1, 2017 and Oct 15, 2017 according to their prior distributions.}
        \label{table:literature}
	\begin{tabular}{lrr} % four columns, alignment for each
		\hline
		 &  $ApJ$ & $MNRAS$\\
		\hline
(a) Jointly proper priors& 18 & \textcolor{black}{34}\\
(b) Jointly improper priors& 1 & \textcolor{black}{2}\\
(c) Unclear priors& 11 & 9 \\		
Total & 30 & 45 \\
                 \hline
	\end{tabular}
\end{table}

\section{Discussion}\label{section4solution}

\subsection{\textcolor{black}{Proving posterior propriety}}\label{sec41}
Improper prior distributions are widely used because they are \textcolor{black}{mathematically} convenient\footnote{We do not consider computational convenience including conjugacy \textcolor{black}{because} most astronomers are familiar with  generic MCMC samplers, such as \texttt{PyStan} \citep{carpenter2017stan}, \texttt{JAGS} \citep{denwood2016jags}, and \texttt{emcee} \citep{foreman2013emcee}. \textcolor{black}{These generic samplers}  automatically sample the target posterior given \textcolor{black}{the likelihood and prior specifications, which enables choosing much wider classes of priors}. \textcolor{black}{\texttt{PyStan} and \texttt{JAGS} always require using proper priors, preventing potential posterior impropriety}.} and \textcolor{black}{are} considered non-informative.  \textcolor{black}{A uniform($-\infty, \infty$)} prior on a location parameter, e.g., $p_1(\theta)$ in Equation~\eqref{improperprior}, \textcolor{black}{is a Jeffreys' prior. It also} has an advantage to make the data (likelihood function) speak more about the parameter when prior knowledge is limited\textcolor{black}{, and results} in a proper posterior distribution in many cases.
However, there is a cost to be paid for using improper priors, which is often neglected:  \emph{Proving posterior propriety}. \textcolor{black}{Adopting an improper prior  for even one parameter requires proving that the integral of a posterior kernel function over the entire parameter space is finite. It is challenging to develop a universal rule-of-thumb about when improper priors are likely to cause improper posterior and when they are not. This is because posterior propriety cannot be assured before it is actually proven on a case by case basis. The problematic choice $d\sigma^2/\sigma^2$ in Section~\ref{section2example}, for example, does not cause posterior impropriety for a different Gaussian model such as $y_j\mid\mu, \sigma^2\sim\textrm{N}(\mu, \sigma^2)$. With $p(\mu, \sigma^2)\propto 1/\sigma^2$, the resulting posterior is proper if $n\ge2$; see Appendix~\ref{appendixD} for a proof.}
\\
\indent There are several ways to prove \textcolor{black}{posterior propriety}.  The most rigorous one is to analytically show that the integral of the target posterior kernel function \textcolor{black}{over the entire parameter space} is finite. However, if the dimensions are large and the model is complicated, \textcolor{black}{which is usually the case in the astronomical literature,} it is challenging to prove posterior propriety analytically. 
\\
\indent\textcolor{black}{We can also} apply  existing theorems about posterior propriety  only if a model considered in a theorem is  the same as  a \textcolor{black}{candidate} model to be used. For example, suppose a \textcolor{black}{candidate} model has two more parameters than a model whose posterior propriety is proven in a theorem. Posterior propriety of \textcolor{black}{the candidate} model holds if a marginalized \textcolor{black}{candidate} model (with  the two additional parameters integrated out from \textcolor{black}{the candidate} model) is the same as \textcolor{black}{the} model considered in the theorem. This is because an unexpected term that is a function of unknown parameters may arise during the integration, which can make  seemingly similar models completely different. 
%A  justification based on conditional densities (or conditional kernel functions) does not necessarily guarantee the finite integral. For example, suppose we are interested in a kernel function,  $q(x, y)\propto \exp(-xy)$ for $x, y>0$. Its impropriety becomes clear once we analytically integrate out \textcolor{black}{both $x$ and $y$}, while its conditional kernel functions, $q(x\mid y)$ and $q(y\mid x)$, are finitely integrable without implying any impropriety \citep{hobert1996propriety}.  

% for some improper priors
%when it is  difficult to prove posterior propriety, we can prevent potential posterior impropriety by 
%The resulting posterior inference with a proper diffuse prior will be similar to the one with \textcolor{black}{uniform$(-\infty, \infty)$} prior. However, the former does not require users to prove posterior propriety while the latter does.} 
%However, the former does not require users to prove posterior propriety while the latter does.
Jointly proper priors  guarantee  posterior propriety based on  standard probability theory. \textcolor{black}{Thus, when researchers want to adopt physically motivated improper priors whose posterior propriety is challenging to be proven, it is a useful practice to adopt proper priors that can mimic the behavior of the improper ones. The resulting posterior inference with mimicking proper priors will be almost identical to the one with improper priors.} For a location parameter whose support is a real line, e.g., $\theta$  in Equation~\eqref{improperprior}, a diffuse Gaussian or diffuse Student's $t$ prior \textcolor{black}{with an arbitrarily large scale can approximate an improper flat prior. The arbitrarily large scale of such a diffuse prior is a computational trick to approximate the improper flat prior although the scale itself may not make sense in practice.}  As for a parameter \textcolor{black}{defined on} a positive real line, e.g., $\sigma^2$ in Equation~\eqref{improperprior}, a log-Normal, half Normal, and half Student's $t$ with relatively large variance are known to  be vague choices \citep{gelman2006prior} \textcolor{black}{that can approximate an  improper flat prior $d\sigma^2$.} \textcolor{black}{Also, a uniform shrinkage prior, $d\sigma^2/(c+\sigma^2)^2$, where $c$ is set to an arbitrarily large constant, can approximate $d\sigma^2$ with good frequentist coverage properties \citep{tak2017usp}.}
%centered at zero 
%Also, proper priors are ideal for constructing scientifically motivated priors 
% but it is too difficult to prove posterior propriety
%scientifically (or physically) motivated 
%scientifically motivated 
%, and   a proof for posterior propriety is not needed
% for some  improper priors

\subsection{\textcolor{black}{Scientifically motivated proper priors for posterior propriety}}\label{sec42}
%daniels1999prior, 
%such as $p_1(\theta)$
%(Improper flat priors for location parameters are scale-free Jeffreys' priors.) 
%textcolor{red}{As mentioned, a} proper prior \textcolor{black}{can be used as an approximation to} a  \textcolor{black}{prior}. 
% a Bayesian model equipped with 
% If both proper and improper priors produce nearly identical posterior distributions, the proper approximation to improper priors is preferred. This is because their similar posterior distributions do not necessarily guarantee posterior propriety of the model with improper priors (even though the resulting inference with improper priors may not be misleading in this case).
%The fact that the posterior with proper priors is almost identical to the one with scientifically motivated improper priors 

%, e.g., $\theta$ in Equation~\eqref{improperprior}
%e.g., $\theta$ in Equation~\eqref{improperprior}, 

%informative, weakly informative \citep{gelman2006prior}, or vague choices. F
%; see also  for other non-informative or weekly informative choices.
%for half-Cauchy and inverse-Gamma priors 
%This  Gaussian prior is still computationally convenient; for example, a defuse Gaussian prior on $\theta$ instead of an improper flat in~\eqref{improperprior} still results in a Gaussian  conditional posterior distribution of $\theta$ as in~\eqref{conditionals} but with slightly different mean and variance.
%-informative) and computationally convenient 

\begin{figure}
\begin{center}
%\plotone{ex1.eps}
\includegraphics[scale = 0.5]{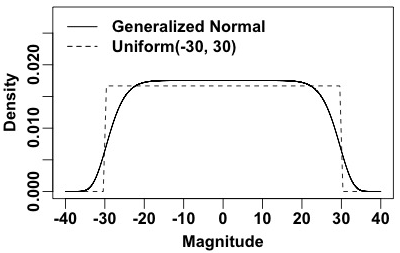}
\caption{\textcolor{black}{Prior distributions of magnitude. The density of the generalized Gaussian distribution with $\mu=0$, $\sigma=30$, and $s=10$ is denoted by a solid curve and that of the uniform($-30, 30$) distribution is represented by a dashed curve.  As $s$ increases, the two densities become close in shape. Unlike the uniform distribution whose hard bounds completely exclude a certain range of magnitude,  the generalized Gaussian distribution sets up soft bounds.} \label{gnorm}}
\end{center}
\end{figure}

%:
%Proper priors can also be used to construct scientifically motivated 
\textcolor{black}{Adopting} scientifically motivated priors is one advantage of using Bayesian machinery because it provides a natural way to incorporate scientific knowledge into inference via priors. Proper priors are ideal for this purpose. \cite{tak2017bayesian},  for example, use  a uniform($-30, 30$) prior for the unknown mean magnitude  of a damped random walk  process, considering a practical magnitude range from that of the Sun to that of the faintest object identifiable by the Hubble Space Telescope. This prior can be considered weakly informative because  the range of the uniform prior is wide enough not to affect the resulting posterior inference. (A bounded uniform prior is not non-informative because its hard bounds completely exclude a certain range of parameter values.) If the range of a uniform prior is narrow and thus it significantly influences the  posterior inference, such an informative choice may need further justification. 
%\textcolor{black}{; see Table~\ref{table:prior} for a summary of a few proper priors that we discusscan easily reflect on scientific knowledge and past studies in .}

\textcolor{black}{If one is uncomfortable about completely excluding a certain parameter space, a generalized Gaussian distribution  \citep{nadarajah2005, nadarajah2006}, also called a power exponential distribution, can be used to set up soft bounds. These soft bounds allow values outside the bounds with small but non-zero probability. Its kernel function of $x$  is proportional to $\exp(-\vert (x-\mu)/\sigma\vert^s)$, where $\mu$ is the location parameter, $\sigma$ is the scale parameter, and $s$ is the shape parameter. The distribution approaches  the uniform($\mu-\sigma, \mu+\sigma$) distribution, i.e.,  the tails of its density decrease more sharply, as $s$ goes to infinity.  Figure~\ref{gnorm} displays its density function for $\mu=0$, $\sigma=30$, and arbitrarily chosen shape parameter $s=10$ with the density of uniform($-30, 30$) superimposed. A generalized Student's $t$ distribution \citep{mcdonald1988gt} can be an alternative if one prefers geometrically decreasing tails so that the data (likelihood) can dominate these bounds more easily (i.e., less informative).}
% as soft bounds
%, which is the weaker prior choice than the generalized Gaussian distribution
% instead of uniform's hard bounds
%If the posterior inference is sensitive to the range of a scientifically motivated Uniform prior, such an informative choice needs further justification. 
%i.e., a small change of the range changes the resulting posterior inference, such  must be further justified.

%Unlike an unbounded improper Uniform prior, a bounded proper one is  an informative choice because of the hard bounds that completely exclude some range of parameter values. The lower and upper bounds, however, can be justified if these account for some scientific knowledge or past studies. 

\begin{figure}
\begin{center}
%\plotone{ex1.eps}
\includegraphics[scale = 0.5]{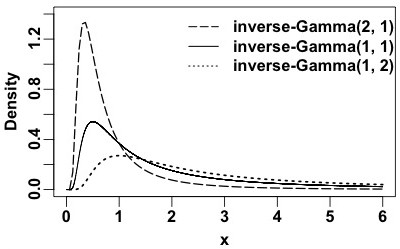}
\caption{\textcolor{black}{Three inverse-Gamma($a, b$) densities according to different shape and scale parameters, $a$ and $b$. We suggest using an inverse-Gamma($a, b$) prior as a way to set up a soft lower bound of a random variable $x$ a priori. The right tail geometrically decreases with power $a+1$ and the left tail exponentially decreases, which indicates that $x$ is less likely to take on values much smaller than the mode (soft lower bound), $b/(a+1)$, a priori. A small value of the shape parameter is desirable for a weakly informative prior. Fixing the shape parameter first ($a\ll n$), we can adjust the scale parameter to set a scientifically motivated soft lower bound}. \label{invgamma}}
\end{center}
\end{figure}

%e relationship between the inverse-Gamma and scaled inverse-$\chi^2$ distributions
% (See, e.g., \cite{gelman2013bayesian} for a discussion of the pseudo observation interpretation of prior distributions.) 
For an unknown parameter whose support is the positive real line, an inverse-Gamma prior can be used as  a scientifically motivated prior because it enables us to set up a soft lower bound of a parameter using scientific knowledge or past studies. The kernel function of  $x$ that follows an inverse-Gamma($a, b$) distribution is $x^{-a-1}\exp(-b/x)$. Its mode,  $b/(a+1)$, plays a role of the soft lower bound, and a small shape parameter $a$ is desirable for a weakly informative prior\footnote{\textcolor{black}{An inverse-Gamma($a, b$) prior is equivalent to an inverse-$\chi^2$ prior with its degrees of freedom $2a$ and scale $b/a$. This relationship}  allows us to interpret the shape parameter of the inverse-Gamma as half the number of pseudo realizations that would carry equivalent information as the prior distribution. For example, an inverse-Gamma prior with the unit shape parameter, $a=1$,  carries a relatively small amount of information from two pseudo observations. If the number of observed data is much larger than two,  the likelihood can dominate this inverse-Gamma prior with ease.}.  When $x$ goes to infinity, the right tail of this kernel function  decreases as a power law,  while the left tail exponentially decreases as $x$ approaches zero. Thus $x$ is less likely to take on values much smaller than the mode (soft lower bound) a priori. \textcolor{black}{See Figure~\ref{invgamma} for a few density curves of inverse-Gamma($a, b$) prior according to different values of $a$ and $b$.} Modeling quasar variability, for example, \cite{tak2017bayesian} adopt an inverse-Gamma(1, $b$) prior for the unknown timescale (in days) of a damped random walk process. The scale parameter $b$ is set to one day  so that  its soft lower bound, 0.5 day, is much smaller than any timescale estimates of 9,275 quasars in a past study \citep{macleod2010modeling} a priori.

%shape parameter determines the The density of the generalized Gaussian distribution with $\mu=0$, $\sigma=30$, and $s=10$ is denoted by a solid curve and that of the Uniform($-30, 30$) distribution is represented by a dashed curve.  As $s$ increases, the two densities become close in shape. Unlike the uniform distribution whose hard bounds completely exclude a certain range of magnitude,  the generalized Gaussian distribution sets up soft bounds.}
For a second-level variance component in a Gaussian hierarchical model such as $\sigma^2$ in Equation~\eqref{normalmodel}, \citet{gelman2006prior} does not recommend  an inverse-Gamma($a, b$) prior with arbitrarily small values \textcolor{black}{for both} $a$ and $b$ as a non-informative choice\footnote{\textcolor{black}{The inverse-Gamma($a, b$) density of $\sigma^2$ behaves similarly to $d\sigma^2/\sigma^2$ as $a$ and $b$ go to zero. However, a difference between two densities is that as $\sigma^2$ goes to zero, the former goes to zero,  while the latter goes to infinity (possibly causing posterior impropriety).}}.  This makes sense because an inverse-Gamma prior always sets up a soft lower bound a priori. When the likelihood puts significant weight at zero but with relatively small data size, it is difficult for the likelihood to dominate the soft lower bound \textcolor{black}{that is located near zero}. In this case, the resulting posterior inference becomes sensitive to the location of the soft lower bound. Thus when the data size is small, it is important to construct the soft lower bound carefully, considering scientific knowledge or past studies. 
%much smaller than the place where most likelihood mass resides
% with relatively small data size  but with small sample size, the soft lower bound may affect the resulting posterior inference.

\begin{table*}
	\centering
	\caption{\textcolor{black}{A few proper priors that can be set up easily to reflect  scientific knowledge and past studies in a weakly informative way.}}
        \label{table:prior}
	\begin{tabular}{cccl} % four columns, alignment for each
		\hline
\textcolor{black}{Distribution} & 	\textcolor{black}{Support}	 &  \textcolor{black}{Kernel function} & \textcolor{black}{Note}\\
		\hline
\textcolor{black}{uniform($a, b$)} & \textcolor{black}{$\mathbb{R}$} &  \textcolor{black}{$1/(b-a)$} & \textcolor{black}{Its hard bounds $(a, b)$, where $a<b~(\in\mathbb{R})$, can reflect  past studies.}\\
\textcolor{black}{generalized Gaussian} & \textcolor{black}{$\mathbb{R}$} & \textcolor{black}{$\exp(-\vert (x-\mu)/\sigma\vert^s)$} & \textcolor{black}{$\mu~(\in\mathbb{R})$ is the location parameter, $\sigma~(\in\mathbb{R}^+)$ is the scale parameter, and}\\
& &  & \textcolor{black}{$s~(\in\mathbb{R}^+)$ is the shape parameter. As $s\to\infty$, this distribution becomes}\\
& &  & \textcolor{black}{uniform($\mu-\sigma, \mu+\sigma$), and thus $(\mu-\sigma, \mu+\sigma)$ can be considered as}\\
& &  &  \textcolor{black}{soft bounds set to represent  scientific knowledge. The choice of $s$ may be}\\
& &  &  \textcolor{black}{arbitrary. A generalized $t$ distribution can be an alternative whose tails}\\
& &  &  \textcolor{black}{decreases geometrically.}\\
\textcolor{black}{inverse-Gamma($a, b$)} & \textcolor{black}{$\mathbb{R}^{+}$}& \textcolor{black}{$x^{-a-1}\exp(-b/x)$} & \textcolor{black}{$a~(\in\mathbb{R}^+)$ is the shape parameter and is treated as the amount of prior}\\
& &  & \textcolor{black}{information. It is desirable to be small for a weakly informative prior}\\
& &  & \textcolor{black}{($a\ll n$). Given $a$, the scale parameter $b~(\in\mathbb{R}^+)$ is set to form a soft lower}\\
& &  & \textcolor{black}{bound $b/(a+1)$ that represent  past studies.}\\
\textcolor{black}{multiply-broken} & \textcolor{black}{$\mathbb{R}^{+}$} & \textcolor{black}{$x^{-\alpha_1}I_{(0<x\le b_1)}\textcolor{black}{+}$} & \textcolor{black}{Small values of powers, $\alpha_1, \ldots, \alpha_k$, are desirable for a weakly informative}\\
\textcolor{black}{power-law} & & $\textcolor{black}{\beta_2}x^{-\alpha_2}I_{(b_1<x\le b_2)}\textcolor{black}{+}$ &  prior. \textcolor{black}{When $k=1$, $\alpha_1\in[0, 1)$ and $b_0=0$. If $k=2$, $\alpha_1\in[0, 1)$ and $\alpha_2>1$.}\\
& & $\cdots  \textcolor{black}{+}$ &  \textcolor{black}{For $k\ge3$, $\alpha_1\in[0, 1)$, $\alpha_j\ge0~\textrm{for}~j=2, 3, \ldots, k-1$, and $\alpha_k>1$.} Segment-\\
& & $\textcolor{black}{\beta_k}x^{-\alpha_k}I_{(b_{k-1}<x)}$&  \textcolor{black}{wise uniform priors are feasible if $\alpha_i=0$ ($i\neq k$). \textcolor{black}{If $\beta_i=b_{i-1}^{\alpha_i - \alpha_{i-1}}$ for $i\ge2$,}}\\
& & & \textcolor{black}{this power law becomes a continuous function}. All powers \textcolor{black}{and coefficients,}\\
& & &  \textcolor{black}{$\alpha_i$, $b_i$, and $\beta_i$,} can reflect scientific knowledge.\\
                 \hline
	\end{tabular}
\end{table*}

\textcolor{black}{A multiply-broken power-law density, proposed by Professor Eric B.~Ford during a personal communication, can be another easy-to-construct scientifically inspired prior for parameters defined on a positive real line. For example, a doubly-broken power-law density is defined as $p(x)\propto x^{-\alpha_1}$ for $0< x\le b_1$, $\textcolor{black}{p(x) = \beta_2x^{-\alpha_2}}$ for $b_1<x\le b_2$, and $\textcolor{black}{p(x) = \beta_3x^{-\alpha_3}}$ for $b_2<x$. \textcolor{black}{If $\beta_i=b_{i-1}^{\alpha_i - \alpha_{i-1}}$ for $i=2, 3$, it becomes a smoothly broken power-law \citep[e.g.,][]{anch2014cosmic}.} Small \textcolor{black}{values of} powers, $\alpha_1 \textcolor{black}{(<1)}$,  $\alpha_2$, and $\alpha_3$ $(>\!1)$, are desirable for weakly informative priors, and zero powers for $\alpha_1$ and $\alpha_2$ enable segment-wise uniform priors. All these parameters including the cut-offs $b_1$ and $b_2$ ($0<b_1<b_2$) need to reflect  astronomical knowledge or past studies.}

%\citep{han2006afterglow} 
%\citep{beuermann1999vlt}
\textcolor{black}{We summarize all these proper priors in Table~\ref{table:prior}.}

\subsection{\textcolor{black}{Re-analysis of the example in Section~\ref{section2example} with jointly proper priors}}
Let us revisit the example in Section~\ref{section2example} to see an impact of adopting jointly proper priors. Instead of the improper choice\textcolor{black}{s} in Equation~\eqref{improperprior}, we set a diffuse Gaussian prior for $\theta$ and a weakly informative inverse-Gamma prior for $\sigma^2$ independently:
\begin{equation}\label{properprior}
\theta\sim \textrm{N}(0,~10^5)~~\textrm{and}~~\sigma^2\sim\textrm{inverse-Gamma}(\textcolor{black}{10^{-2}, 1}).
\end{equation}
\textcolor{black}{We first set the shape parameter of the inverse-Gamma($a, b$) prior to $10^{-2}$ that is much smaller than the data size ($a\ll n$). Next we set $b=1$ to construct a soft lower bound, 0.99, assuming that it reflects  scientific knowledge a priori.} We denote the joint prior distribution in Equation~\eqref{properprior}  by $p^\ast(\theta, \sigma^2)$. The resulting  full posterior kernel function is
\begin{equation}\label{properpost}
q^\ast(\boldsymbol{\mu}, \theta, \sigma^2 \mid \boldsymbol{y}) = p^\ast(\theta, \sigma^2)\prod_{j=1}^{n}\left[f(y_j\mid \mu_j)~ p(\mu_j\mid \theta, \sigma^2)\right],
\end{equation}
where density functions $f$ and $p$ are defined in Equation~\eqref{normalmodel}. The corresponding Gibbs sampler updates each coordinate of $\boldsymbol{\mu}$ by its conditional posterior specified in Equation~\eqref{conditionals} but updates $\theta$ and $\sigma^2$ by
\begin{align}
\begin{aligned}\label{conditionals2}
\theta\mid \boldsymbol{\mu}, \sigma^2,  \boldsymbol{y} & \sim\textrm{N}\left(~\frac{(n/\sigma^2)\bar{\mu}}{n/\sigma^2+1/10^5},~ \frac{1}{n/\sigma^2+1/10^5}~\right),\\
\sigma^2\mid \boldsymbol{\mu}, \theta,  \boldsymbol{y} & \\
\sim\textrm{inverse}&\textrm{-Gamma}\left(~\frac{n}{2}+\frac{1}{10^2},~ \frac{\sum_{j=1}^n(\mu_j-\theta)^2}{n}+1~\right).
\end{aligned}
\end{align}
\textcolor{black}{The conditional distribution of $\theta$ in Equation~\eqref{conditionals2} is similar to that in Equation~\eqref{conditionals}, considering that  $10^{-5}$ is} close to zero. The other simulation configuration is the same.

%while we do not know whether the posterior sample displayed in Figure~\ref{ex1figure} comes from $q$ in equation~\eqref{improperpost}
%, which again emphasizes the importance of checking posterior propriety before Bayesian analyses

\begin{figure*}
\begin{center}
%\plotone{ex1.eps}
\includegraphics[scale = 0.5]{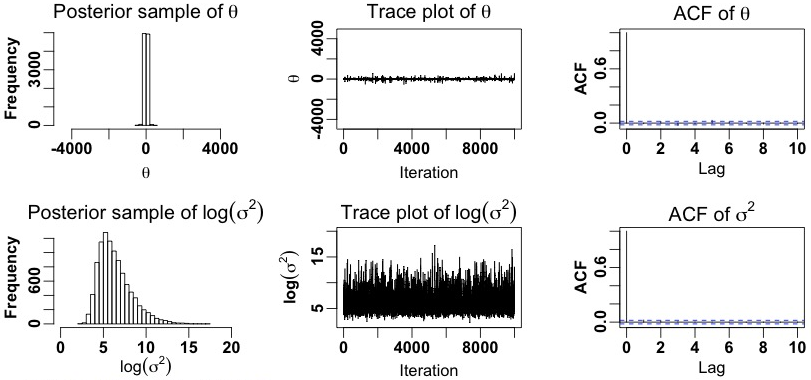}
\caption{The result of sampling the proper posterior kernel function in Equation~\eqref{properpost} that is based on weakly informative and vague proper priors in Equation~\eqref{properprior}. The histogram, trace plot, and auto-correlation function of 10,000 posterior samples of $\theta$ are on the top  panels and those of $\log(\sigma^2)$ on the bottom panels. The ranges of vertical and horizontal axes are the same as those in Figure~\ref{ex1figure}. Since priors in Equation~\eqref{properprior} are jointly proper,  we know that the resulting posterior  $q^\ast$ in Equation~\eqref{properpost} is proper and thus the posterior sample is from $q^\ast$. Also, the sampling result hardly varies even if the initial value of $\sigma^2$ is close to zero. The ensuing Bayesian inference on $\theta$ is  quite different from that in Section~\ref{section2example} with much shorter tails. \label{ex2figure}}
\end{center}
\end{figure*}

Figure~\ref{ex2figure} exhibits the sampling result. The ranges of the horizontal and vertical axes in each panel are the same as those of Figure~\ref{ex1figure} for a  comparison. Because of the jointly proper priors in Equation~\eqref{properprior},  we know that the resulting posterior kernel function $q^\ast$ in Equation~\eqref{properpost} is proper and thus the posterior sample displayed in Figure~\ref{ex2figure} represents the target posterior distribution.  Although not shown here, the MCMC method produces nearly the same sampling result regardless of the initial value of $\sigma^2$\textcolor{black}{, meaning that the Markov chain converges no matter where it starts.} The histogram of $\theta$ in Figure~\ref{ex2figure} has much shorter tails than that in Figure~\ref{ex1figure}, although the histogram of $\log(\sigma^2)$ in Figure~\ref{ex2figure} is similar to that in Figure~\ref{ex1figure}. The soft lower bound of $\log(\sigma^2)$, i.e., \textcolor{black}{$\log(0.99)=-0.01$, is not close to the high density region}. This indicates that \textcolor{black}{the} soft lower bound does not affect the \textcolor{black}{resulting} posterior inference  even though there are just two data points; the degrees of freedom \textcolor{black}{parameter} of an equivalent inverse-$\chi^2$ prior \textcolor{black}{is 0.02}. \textcolor{black}{A sensitivity analysis, though not shown here, indicates that the results are robust as long as the scale parameter $b$ of the inverse-Gamma($10^{-2}, b$) prior puts the soft lower bound on the left-hand side of the high-density region.} The effective sample size improves greatly; it is \textcolor{black}{9,823} for $\theta$ and \textcolor{black}{2,668} for $\log(\sigma^2)$. Consequently, the inference on $\theta$ becomes quite different from that in Section~\ref{section2example}, empirically \textcolor{black}{showing} that checking posterior propriety before using MCMC methods can make a significant difference. 
%
%  On the other hand, we do not know where the posterior sample displayed in Figure~\ref{ex1figure} comes from. 
%that this inverse-Gamma prior is not informative 
%If We clearly know that the posterior sample shown in Figure~\ref{ex2figure} comes from the target posterior distribution while that in Figure~\ref{ex1figure} does not.

%Proving posterior propriety requires specifying a Bayesian model, i.e., a likelihood function of unknown parameters and their prior distributions. 
%\textcolor{black}{Specifying the Bayesian model in appendix will be helpful for both readability and reproducible science.} 
%finitely integrable
%, and  the resulting inference may or may not be based on a probability distribution
%Posterior propriety is fundamental in Bayesian analysis, satisfying one of two conditions that make a posterior distribution a probability distribution and two of three conditions that guarantee Markov chain's unique stationary distribution. 
%result in proper posterior distributions. 

\subsection{\textcolor{black}{Concluding remarks}}

It is well understood that any probabilistic \textcolor{black}{tool} such as Bayesian \textcolor{black}{interface} should be based on a probability distribution. Jointly proper priors lead to a proper posterior distribution and can be either vague or scientifically motivated.  However,  improper priors are sometimes used to represent the lack of prior knowledge. Such improper priors combined with a likelihood function can result in an improper posterior distribution that is not \textcolor{black}{a probability measure}. Therefore, when improper priors are adopted, posterior propriety must be carefully proven before using any MCMC methods\textcolor{black}{, which can also improve statistical clarity and reproducibility}. We hope that posterior propriety draws more attention  when improper priors are used in the astronomical literature.
\\
\indent\textcolor{black}{For a more complete Bayesian analysis, posterior propriety is not the only thing to be checked in practice. We list some procedures or practices essential for a good Bayesian analysis as the referee recommends. Above all, we re-emphasize the importance of clarifying a Bayesian model, i.e., clearly specifying both likelihood function and priors, which is the beginning of the Bayesian analysis that  precedes even checking posterior propriety. During the analysis, it is  important to explore the entire (pre-specified) parameter space by implementing multiple Markov chains whose initial values are spread across the parameter space; it is desirable to specify the initial values of the chains. After the analysis, a good Bayesian analysis comes with various diagnostic procedures. An MCMC convergence check is necessary in both visual and numerical ways, e.g., trace plots, auto-correlation functions, effective sample sizes, and Gelman-Rubin diagnostic statistics \citep{gelman1992inference}. As for model checking \citep[Section~6,][]{gelman2013bayesian}, a posterior predictive check is a valuable tool to assess a model's consistency with the observed data, i.e., whether the model can explain the data generation process well.  A prior predictive check (if priors are proper) is also useful for  model checking, which generates a data set using a model with known parameter values and checks whether a fitted model can recover the generative parameter values. In addition, a sensitivity check is  important to understand the influence of  prior assumptions on the resulting posterior inference. We hope good Bayesian practice becomes more popular in the astronomical literature for more reliable Bayesian analysis.}
%A full investigation about these issues is already well documented in  statistical literature, but may be challenging in the astronomical literature. 
% investigation is unnecessary 
%This is because a model is a set of assumptions, and thus results from a model are valid only in the context of assumptions. 
% as future research for a more complete Bayesian analysis in the astronomical literature
% Without a clearly specified model, the results may be This. Without this, researchers would present scientific findings If readability becomes a concern,
%There are more things to do other than checking posterior propriety
% and there are more things to do to the Bayesian analysis in a complete way

\section*{Acknowledgements}

Hyungsuk Tak and Sujit K.~Ghosh acknowledge partial support from  National Science Foundation grant DMS 1127914 (and DMS 1638521 only for Hyungsuk Tak) given to the  Statistical and Applied Mathematical Sciences Institute.  Justin A.~Ellis acknowledges supports from the National Science Foundation Physics Frontier Center Grant 1430284 and from the National Aeronautics and Space Administration through Einstein Fellowship Grant PF4-150120. We thank David E.~Jones and David C.~Stenning for a productive discussion at the International Centre for Theoretical Sciences in Bangalore, India, during a visit when participating in the program `Time Series Analysis for Synoptic Surveys and Gravitational Wave Astronomy'. \textcolor{black}{We also thank Eric B.~Ford for insightful comments,  Christian P.~Robert and Peter Coles for their discussions in their personal blogs, and the referee for invaluable suggestions.}
%%%%%%%%%%%%%%%%%%%%%%%%%%%%%%%%%%%%%%%%%%%%%%%%%%



%\bibliographystyle{mnras}


%%%%%%%%%%%%%%%%%%%%%%%%%%%%%%%%%%%%%%%%%%%%%%%%%%

%%%%%%%%%%%%%%%%% APPENDICES %%%%%%%%%%%%%%%%%%%%%

\appendix

\section{The Bayes Theorem in Detail}\label{appendixBayes}

It is well known that a Bayesian statistical model consists of (i) a sampling distribution, $f(\boldsymbol{y}\mid\boldsymbol{\theta})$, denoting the conditional probability density of the data $y$ given unknown parameters $\boldsymbol{\theta}$; and (ii) a prior distribution, $p(\boldsymbol{\theta})$, denoting an unconditional probability density of $\boldsymbol{\theta}$.  The resulting joint density of $\boldsymbol{y}$ and $\boldsymbol{\theta}$ is $f(\boldsymbol{y}\mid\boldsymbol{\theta})p(\boldsymbol{\theta})$ based on standard probability theory. We can also  express this joint density as a product of the unconditional density of the data $h(\boldsymbol{y})\equiv\int f(\boldsymbol{y}\mid\boldsymbol{\theta})p(\boldsymbol{\theta})d\boldsymbol{\theta}$ and the so-called posterior density of $\boldsymbol{\theta}$ given $\boldsymbol{y}$, i.e., 
\begin{equation}
\pi(\boldsymbol{\theta}\mid \boldsymbol{y})=\frac{f(\boldsymbol{y}\mid\boldsymbol{\theta})p(\boldsymbol{\theta})}{\int f(\boldsymbol{y}\mid\boldsymbol{\theta})p(\boldsymbol{\theta})d\boldsymbol{\theta}}.
\label{bthm}
\end{equation}
All density functions are formally defined with respect to  Lebesgue measure (or counting measure). However, in many scientific applications, we may relax the need for the use of a probability measure for the prior distribution by using a kernel function $k(\boldsymbol{\theta})=c_0p(\boldsymbol{\theta})$ for some constant $c_0>0$ and also write the likelihood function $L(\boldsymbol{\theta}; \boldsymbol{y})=c(\boldsymbol{y})f(\boldsymbol{y}\mid\boldsymbol{\theta})$ for some function $c(\boldsymbol{y})>0$. Then, as illustrated in \citet{ghosh2010basics}, we can reexpress  Equation~\eqref{bthm} as 
\begin{equation}
\pi(\boldsymbol{\theta}\mid \boldsymbol{y})=\frac{f(\boldsymbol{y}\mid\boldsymbol{\theta})p(\boldsymbol{\theta})}{\int f(\boldsymbol{y}\mid\boldsymbol{\theta})p(\boldsymbol{\theta})d\boldsymbol{\theta}}= \frac{L(\boldsymbol{\theta}; \boldsymbol{y})k(\boldsymbol{\theta})}{\int L(\boldsymbol{\theta}; \boldsymbol{y})k(\boldsymbol{\theta})d\boldsymbol{\theta}}.
\label{bthm2}
\end{equation}
As illustrated in Section~\ref{sec:intro}, even if $\int k(\boldsymbol{\theta})d\boldsymbol{\theta}=\infty$, making the prior distribution improper, the posterior density as given in Equation~\eqref{bthm2} is still a valid probability density as long as the denominator $\int L(\boldsymbol{\theta}; \boldsymbol{y})k(\boldsymbol{\theta})d\boldsymbol{\theta}<\infty$ is finitely integrable. However, an improper prior necessarily leads to improper marginal distribution of the data $\boldsymbol{y}$ (and vice versa), i.e., $\int k(\boldsymbol{\theta}) d\boldsymbol{\theta}=\infty$ is equivalent to $\int h(\boldsymbol{y}) d\boldsymbol{y}=\infty$. This is because 
\begin{align}
\int h(\boldsymbol{y}) d\boldsymbol{y}&=\int\int f(\boldsymbol{y}\mid\boldsymbol{\theta})p(\boldsymbol{\theta})d\boldsymbol{\theta}~\! dy\nonumber\\
&=\int\int f(\boldsymbol{y}\mid\boldsymbol{\theta})p(\boldsymbol{\theta})d\boldsymbol{y}~\!d\boldsymbol{\theta}=\int p(\boldsymbol{\theta}) d\boldsymbol{\theta},\nonumber
\end{align}
where the second equality holds from Fubini's theorem. This aspect is not a concern  if $\pi(\boldsymbol{\theta}\mid \boldsymbol{y})$ is a proper probability density. It is  well  known that, in order to make an  inference about $\boldsymbol{\theta}$ (or its function) conditional on the observed data, it is often sufficient to draw samples from a posterior kernel given by $L(\boldsymbol{\theta}; \boldsymbol{y})k(\boldsymbol{\theta})$, i.e., the numerator in Equation~\eqref{bthm2} without the need to evaluate the denominator.  Unfortunately, Gibbs-type MCMC methods can generate a sample from the posterior kernel which need not correspond to a proper posterior distribution; see \citet{hobert1996propriety} for various examples. When a proper prior density $p(\boldsymbol{\theta})$ is used, this is not an issue as a posterior distribution is necessarily proper by  standard probability theory. However, when an improper prior kernel  is used, then the only option is to verify \emph{analytically} that integral in the denominator of Equation~\eqref{bthm2} is finite.
%\textcolor{black}{(Sujit, I removed ABC because it may require using proper priors. Tak: sounds fine)}
%$k(\boldsymbol{\theta})$
%a posterior probabilistic inference is based on the conditional distribution of $\boldsymbol{\theta}$ given $\boldsymbol{y}$
%the posterior distribution which can often be accomplished by using only 

\section{Proof of Posterior Impropriety in Section~2}\label{appendixB}
The full posterior kernel function $q(\boldsymbol{\mu}, \theta, \sigma^2\mid \boldsymbol{y})$ in Equation~\eqref{improperpost} is improper because the marginal posterior kernel function $q_1(\sigma^2\mid \boldsymbol{y})$ with $\boldsymbol{\mu}$ and $\theta$ integrated out from $q(\boldsymbol{\mu}, \theta, \sigma^2\mid \boldsymbol{y})$  is improper. We derive the marginal posterior kernel function of $\theta$ and $\sigma^2$ by integrating out  $\boldsymbol{\mu}$ from  $q(\boldsymbol{\mu}, \theta, \sigma^2\mid \boldsymbol{y})$:
\begin{align}
\begin{aligned}\label{marginal2}
q_2(\theta, &~\sigma^2\mid \boldsymbol{y}) =\int_{\mathbb{R}^{n}}q(\boldsymbol{\mu}, \theta, \sigma^2\mid \boldsymbol{y})~d\boldsymbol{\mu}\\
&=p_1(\theta, \sigma^2)\prod_{j=1}^n f_1(y_j\mid \theta, V_j+\sigma^2)\\
&=\frac{1}{\sigma^2}\exp\left(-\sum_{j=1}^n\frac{(y_j-\theta)^2}{2(V_j+\sigma^2)}  \right)\prod_{j=1}^n (V_j+\sigma^2)^{-0.5}\\
&=\frac{1}{\sigma^2}\exp\left(-\sum_{j=1}^n\frac{(y_j-\hat{y})^2}{2(V_j+\sigma^2)} -\frac{(\theta-\hat{y})^2}{2V^\ast} \right)\prod_{j=1}^n (V_j+\sigma^2)^{-0.5},
\end{aligned}
\end{align}
where density functions $p_1$ and $f_1$ are defined in Equations~\eqref{improperprior} and \eqref{marginaldata}, respectively,
\begin{equation}\nonumber
\hat{y}\equiv \frac{\sum_{j=1}^n y_j /(V_j+\sigma^2)}{\sum_{j=1}^n 1/(V_j+\sigma^2)}~~\textrm{and}~~V^\ast\equiv\frac{1}{\sum_{j=1}^n 1/(V_j+\sigma^2)}.
\end{equation}
Next we marginalize out $\theta$ from Equation~\eqref{marginal2} as follows:
\begin{align}\nonumber
q_1(\sigma^2\mid \boldsymbol{y}) &=\int_{\mathbb{R}}q_2(\theta, \sigma^2\mid \boldsymbol{y})~d\theta\\
&=\frac{(V^\ast)^{0.5}}{\sigma^2}\exp\left(-\sum_{j=1}^n\frac{(y_j-\hat{y})^2}{2(V_j+\sigma^2)}\right)\prod_{j=1}^n (V_j+\sigma^2)^{-0.5}.
\end{align}
This marginal posterior kernel function of $\sigma^2$ approaches infinity as $\sigma^2$ goes to zero due to the prior on $\sigma^2$, i.e., $d\sigma^2/\sigma^2$. Therefore,  $\int_{\mathbb{R}^+}q_1(\sigma^2\mid \boldsymbol{y})~d\sigma^2=\infty$.

\section{Classification of 75 articles in Section~3} \label{appendixC}

\textcolor{black}{On the webpage of IOPscience, we found 33 \emph{ApJ} articles whose titles or abstracts contain a word `Bayesian'. We excluded three of them because one is an erratum \citep{Eadie2017erratum} and the other two  use just Bayesian methods previously developed by other researchers \citep{abeysekara2017daily, murphy2017the}. We also obtained a list of 51 articles from the webpage of \emph{MNRAS} that have the word `Bayesian' in their abstracts. We did not consider six of them  because one mentions a Bayesian analysis as a potential application \citep{watkinson2017a}, another uses a Bayesian information criterion for a model selection \citep{Wilkinson2017firefly}, and the other four simply utilize Bayesian methods  developed in other articles \citep{pinamonti2017searching, green2017the, sampedro2017a, basak2017surprise}}. 

Among the 30 articles published online in \emph{ApJ}, 18 articles adopt jointly proper priors, and we classify these into category (a); \cite{fogarty2017the}; \cite{montes2017comparison}; \cite{zevin2017constraining}; \cite{leung2017bayesian}; \cite{farnes2017observed}; \cite{benson2017mc2}; \cite{sathyanarayana2017modeling}; \cite{sliwa2017luminous}; \cite{park2017extending}; \cite{Khrykin2017the}; \cite{budavari2017faint}; \cite{Wang2017the}; \cite{Scherrer2017a}; \cite{Tabatabaei2017the}; \cite{Lund2017standing}; \cite{Eadie2017bayesian}; \cite{0004-637X-835-1-93}; \cite{0004-637X-834-2-112}.

\cite{kne2017balmer} set an unbounded flat prior on the logarithm of the total flux without proving posterior propriety, and thus we classify this article into category~(b). 
%Posterior propriety in this article is not proven.
%One article published in \emph{ApJ}  uses  improper priors for some parameters, and 

We cannot check posterior propriety of 11 articles published online in \emph{ApJ} because they do not specify priors clearly, i.e., their Bayesian models are not uniquely determined. We designate them as category (c) which contains cases where uniform (or flat) priors are used without clear ranges. Here we list them; \cite{kern2017emulating} use  flat priors over the astrophysical parameters; \cite{bitsakis2017anovel}  say nothing about priors; \cite{raithel2017from} do not specify a joint prior on five pressures; \cite{oyar2017a} utilize  flat priors on  all parameters; \cite{tanaka2017resolved} make use  of  uniform priors on $m_{\textrm{TRGB}}$ and $a$; \cite{mandel2017} adopt a flat prior on $\mu_s$ whose range is unclear;  \cite{Daylan2017inference} adopt  uniform priors on many parameters; \cite{Warren2017sparse}  utilize an uninformative prior on $\beta$; \cite{0004-sp-836-1-43} make use of an uninformative prior on $h$; \cite{Eilers2017joint} do not specify priors on $\gamma$, $\sigma_C$, and $\sigma_{ij}$; and \cite{jones2017measuring} use flat priors on SN Ia  distances. 

%\cite{abeysekara2017daily} and  \cite{murphy2017the} take advantage of the Bayesian block approach \citep{scagle2013} without checking its posterior propriety; 

Next, we classify 45 articles published online in \emph{MNRAS} into three categories. Category (a) contains \textcolor{black}{34} articles whose priors are jointly proper; \cite{ashton2017on}; \cite{bainbridge2017artificial}; \cite{ata2017the}; \cite{wagner-kaiser2017the}; \cite{cibirka2017codex}; \cite{patel2017orbits}; \cite{si2017a}; \cite{dwelly2017spiders}; \cite{maund2017the}; \cite{davis2017wisdom}; \cite{hahn2017approximate}; \cite{burgess2017the}; \cite{silburt2017resonant}; \cite{macDonald2017hd}; \cite{abdurrouf2017understanding}; \cite{kafle2017galactic}; \cite{aigrain2017robust}; \cite{henderson2017a}; \cite{kimura2017rapid}; \cite{schellenberger2017hicosmo}; \cite{Mejia-Narvaez2017galaxy}; \cite{kohlinger2017kids}; \cite{dam2017apparent}; \cite{garnett2017detecting}; \cite{andrews2017wide}; \cite{kovalenko2017maximum}; \cite{McEwen2017wavelet}; \cite{Oh20172d}; \cite{duncan2017photometric}; \cite{galvin2017the}; \cite{Salvato2017finding}; \cite{yu2017on};  \cite{greig2017simultaneously}; \textcolor{black}{and \cite{sereno2017comalit}}\footnote{\textcolor{black}{In earlier preprints of this manuscript, we put their work into category~(b). This is because Table~1 of \cite{sereno2017comalit}  sets \texttt{Z.max} $=+\infty$, resulting in an improper uniform prior on \texttt{mu.Z.0} whose upper limit is infinity. Specifically, the} LIRA manual \citep{sereno2017lira} says, ``\texttt{Z.max}: maximum value of the \texttt{Z} distribution. The Gaussian distribution and the prior on \texttt{mu.Z.0} are truncated above \texttt{Z.max}. If \texttt{n.mixture}>1, \texttt{Z.max} is automatically set to \texttt{n.large}.'' Since the prior distribution on \texttt{mu.Z.0} is a uniform distribution as specified in Table~1 of \cite{sereno2017comalit}, its upper bound is infinity by specifying \texttt{Z.max} $=+\infty$. \textcolor{black}{Although the prior on \texttt{mu.Z.0} is specified as an improper uniform prior in Table~1 of \cite{sereno2017comalit}, their code implementation is based on a bounded uniform prior on \texttt{mu.Z.0} \citep{sereno2017lira, sereno2017improper}. Considering that their reported results are based on their code implementation with jointly proper prior distributions, we now put their work into category~(a) despite the inconsistency between prior specification and code implementation. We hope that in the future the priors on both \texttt{Z} and \texttt{mu.Z.0} are separately specified with clear bounds of the uniform prior on \texttt{mu.Z.0} in a published article for a consistency between prior specification and code implementation.}}.
%
% \textcolor{black}{It is always desirable to avoid such inconsistency between the prior specification and code implementation.}

\textcolor{black}{Two} articles published online in \emph{MNRAS} employ improper priors without proving posterior propriety; \cite{kos2017spatial} sets an improper prior on $l_{\textrm{se}}$  without an upper limit; and we proved  posterior impropriety of \cite{pihajoki2017a} resulting from the improper prior on $\sigma^2$.
% in Section~\ref{section2example}

We cannot judge posterior propriety of 9 articles  published in \emph{MNRAS} because their priors are not clearly specified; \cite{rodrigues2017determining} adopt flat priors on metallicity and age; \cite{vallisneri2017taming} do not specify priors on $\sigma_{\textrm{out}}$ and $c$; \cite{binney2017modeling} use uniform priors for the logarithm of scale parameters; \cite{ashworth2017exploring} use flat priors on $\alpha_3$ and $A_V$; \cite{jeffreson2017the} utilize uniform priors on eight parameters (three are on the logarithmic scale); \cite{molino2017clash} adopt flat priors on galaxy type and redshift; \cite{accurso2017deriving} do not specify priors on $\alpha$ and $\beta_j$; \cite{Gunther2017centroid} adopt uniform priors on all parameters; and \cite{igoshev2017how} do not clarify a joint prior on $\vec{v}_o$.
%\cite{pinamonti2017searching} utilize Bayesian Lomb-Scargle and Bayesian generalized Lomb-Scargle periodograms \citep{mortier2015} without details; 
%\cite{green2017the} adopt a parallax-based Bayesian probability approach \citep{reid2016a} without details; 
%\cite{sampedro2017a} make use of a Bayesian non-parametric method \citep{cabrera1990} without details; 
%\cite{basak2017surprise} use a Bayesian block method \citep{scagle2013} without details; 

\section{Proof of Posterior Propriety in Section~4}\label{appendixD}
\textcolor{black}{The target posterior kernel function of $\mu$ and $\sigma^2$ is as follows:
\begin{equation}\label{d1}
\pi(\mu, \sigma^2\mid \boldsymbol{y})\propto (\sigma^2)^{n/2 - 1}\exp\left[-\frac{\sum_{i=1}^n (y_i-\mu)^2}{2\sigma^2}\right].
\end{equation}
The marginal posterior kernel function of $\sigma^2$ with $\mu$ integrated out from Equation~\eqref{d1} is
\begin{equation}\label{d2}
\pi(\sigma^2\mid \boldsymbol{y})\propto (\sigma^2)^{-(n-1)/2 - 1}\exp\left[-\frac{\sum_{i=1}^n (y_i-\bar{y})^2}{2\sigma^2}\right].
\end{equation}
The integral of $\pi(\sigma^2\mid \boldsymbol{y})$ in Equation~\eqref{d2} is finite if $n$ is greater than 1 because the marginal posterior distribution of $\sigma^2$  is inverse-Gamma($(n-1)/2,~\sum_{i=1}^n (y_i-\bar{y})^2/2$), considering the functional form of $\pi(\sigma^2\mid \boldsymbol{y})$ in Equation~\eqref{d2}.}
%\newpage
\bibliography{bibliography}
\bibliographystyle{apalike}

%%%%%%%%%%%%%%%%%%%%%%%%%%%%%%%%%%%%%%%%%%%%%%%%%%

% Don't change these lines
\bsp	% typesetting comment
\label{lastpage}
\end{document}